\documentclass[twocolumn,showpacs,preprintnumbers,superscriptaddress,amsmath,amssymb]{revtex4}
\usepackage{graphicx}  
\usepackage{epstopdf}
\usepackage{dcolumn}   
\usepackage{bm}        
\usepackage{hyperref}

\newcommand{\columbia}{\affiliation{Physics Department, Columbia University, New York, NY 10027, USA}}
\newcommand{\coimbra}{\affiliation{Department of Physics, University of Coimbra, R. Larga, 3004-516, Coimbra, Portugal}}
\newcommand{\rice}{\affiliation{Department of Physics, Rice University, Houston, TX 77005 - 1892, USA}}
\newcommand{\lngs}{\affiliation{INFN, Laboratori Nazionali del Gran Sasso, Assergi, 67100, Italy}}
\newcommand{\ucla}{\affiliation{Physics and Astronomy Department, University of California, Los Angeles, USA}}
\newcommand{\muenster}{\affiliation{Institut f\"ur Kernphysik, Westf\"alische Wilhelms-Universit\"at M\"unster, 48149 M\"unster, Germany}}
\newcommand{\mainz}{\affiliation{Institute of Physics, Johannes Gutenberg University Mainz, 55128 Mainz, Germany}}
\newcommand{\sjtu}{\affiliation{Department of Physics, Shanghai Jiao Tong University, Shanghai, 200240, China}}
\newcommand{\subatech}{\affiliation{SUBATECH, Ecole des Mines de Nantes, Universit\'e de Nantes, CNRS/IN2P3, Nantes, France}}
\newcommand{\uzh}{\affiliation{Physics Institute, University of Z\"urich, Winterthurerstrasse 190, CH-8057, Z\"urich, Switzerland}}
\newcommand{\heidelberg}{\affiliation{Max Planck Institute for Nuclear Physics, Heidelberg, Germany}}

\begin{document}

\title{Study of the electromagnetic background in the XENON100 experiment}

\author{E.~Aprile}\columbia
\author{K.~Arisaka}\ucla
\author{F.~Arneodo}\lngs
\author{A.~Askin}\uzh
\author{L.~Baudis}\uzh
\author{A.~Behrens}\uzh
\author{K.~Bokeloh}\muenster
\author{E.~Brown}\ucla\muenster
\author{J.M.R.~Cardoso}\coimbra
\author{B.~Choi}\columbia
\author{D.~Cline}\ucla
\author{S.~Fattori}\lngs\mainz
\author{A.D.~Ferella}\uzh
\author{K.-L.~Giboni}\columbia
\author{A.~Kish}\email[Corresponding author. E-mail: ]{alexkish@physik.uzh.ch}\uzh
\author{C.W.~Lam}\ucla
\author{J.~Lamblin}\subatech
\author{R.F.~Lang}\columbia
\author{K.E.~Lim}\columbia
\author{Q.~Lin}\sjtu
\author{S.~Lindemann}\heidelberg
\author{M.~Lindner}\heidelberg
\author{J.A.M.~Lopes}\coimbra
\author{K.~Lung}\ucla
\author{T.~Marrod\'an Undagoitia}\uzh
\author{Y.~Mei}\rice
\author{A.J.~Melgarejo Fernandez}\columbia
\author{K.~Ni}\sjtu
\author{U.~Oberlack}\mainz\rice
\author{S.E.A.~Orrigo}\coimbra
\author{E.~Pantic}\ucla
\author{G.~Plante}\columbia
\author{A.C.C.~Ribeiro}\coimbra
\author{R.~Santorelli}\uzh
\author{J.M.F. dos Santos}\coimbra
\author{M.~Schumann}\uzh\rice
\author{P.~Shagin}\rice
\author{H.~Simgen}\heidelberg
\author{A.~Teymourian}\ucla
\author{D.~Thers}\subatech
\author{E.~Tziaferi}\uzh
\author{H.~Wang}\ucla
\author{M.~Weber}\heidelberg
\author{C.~Weinheimer}\muenster

\collaboration{XENON100 Collaboration}\noaffiliation

\begin{abstract}
The XENON100 experiment, located at the Laboratori Nazionali del Gran Sasso (LNGS), aims to directly detect dark matter in the form of Weakly Interacting Massive Particles (WIMPs) via their elastic scattering off xenon nuclei. We present a comprehensive study of the predicted electronic recoil background coming from  radioactive decays inside the detector and shield materials, and intrinsic radioactivity in the liquid xenon. 
Based on GEANT4 Monte Carlo simulations using a detailed geometry together with the measured radioactivity of all detector components, we predict an electronic recoil background in the WIMP-search energy range and 30~kg fiducial mass of less than 10$^{-2}$~events$\cdot$kg$^{-1}\cdot$day$^{-1}\cdot$keV$^{-1}$, consistent with the experiment's design goal. The predicted background spectrum is in very good agreement with the data taken during the commissioning of the detector in Fall~2009.
\end{abstract}

\keywords{Dark Matter, Xenon, Background, Monte Carlo Simulation, GEANT4} 
\pacs{95.35.+d, 29.40.-n, 34.80.Dp}

\maketitle

\section{Introduction}
\label{intro}

For all experiments dealing with very low signal rates, such as dark matter
or double beta decay searches, the reduction and discrimination of the background is one of the most
important and difficult tasks. As the sensitivity of these experiments
keeps increasing, the fight against the background remains crucial.

The XENON100 detector, which is installed in the  Laboratori Nazionali del Gran Sasso (LNGS), Italy, is the second generation detector within the XENON program, dedicated to the direct detection of particle dark matter in the form of Weakly Interacting Massive Particles (WIMPs)~\cite{wimps}.
It is the successor of XENON10~\cite{xe10-instrument}, which has set some of the best limits on WIMP-nucleon scattering cross-sections~\cite{xe10-independent, xe10-dependent}. XENON100 aims to improve this sensitivity due to an increase of the target mass and a significant reduction of the background in the target volume.

In the standard scenario, WIMPs are expected to elastically scatter off xenon nuclei resulting in low energy nuclear recoils. Neutrons passing through the detector also produce nuclear recoils of similar energy, whereas gamma rays and electrons produce electronic recoils. This opens the possibility to efficiently reject the electromagnetic background using various discrimination techniques. 
Experiments like XENON100 distinguish electronic interactions from nuclear recoils based on a different ratio in the yield of scintillation light (primary signal, S1) and ionization charge (secondary signal, S2). Using this discrimination technique, XENON10 and XENON100 reached an electronic recoil rejection efficiency better than 99\% at 50\% nuclear recoil acceptance~\cite{xe100-independent, xe10-independent}.

The main sources of electronic recoil background in XENON100 are radioactive contamination of the materials used to construct the detector and the shield, intrinsic radioactivity in the LXe target,  
and the decays of $^{222}$Rn and its progeny inside the detector shield. 
Even if the electronic recoil rejection efficiency, based on the ratio of the scintillation and ionization signals, is high, a potential statistical leakage of electronic recoil events into the nuclear recoil region can mimic a dark matter signal. 
One way to handle this background is a well-planned detector design avoiding the presence of radioactive materials close to the active volume.

During the design phase of XENON100, all detector materials and components have been carefully selected based on measurements of their radioactive contamination in order to achieve a low background level. The background is further suppressed by improvements on the passive shield and by surrounding the target volume with an active LXe veto layer.

In this paper we summarize the effort to use extensive Monte Carlo simulations to predict the electronic recoil background of XENON100 from natural radioactivity in the detector and shield components, and to study the background reduction by applying fiducial volume and veto coincidence cuts.
Section~\ref{secDetectorModel} describes the detector model which has been used in the simulations. 
The predicted electronic recoil background from the detector and shield materials is discussed in Section~\ref{secDetectorMaterials}, the background from the decays of $^{222}$Rn and its progenies in the shield cavity in Section~\ref{secRadonCavity}, and the background from radon and krypton in LXe in Section~\ref{secIntrinsicBG}. The comparison of the background model with the measured background spectrum is presented in Section~\ref{secComparison}, and conclusions are drawn in Section~\ref{secConclusions}.

\section{XENON100 detector design and model simulated with the GEANT4 toolkit}
\label{secDetectorModel}

The XENON100 detector is a dual phase time-projection chamber (TPC). The total amount of 161~kg of LXe is enclosed in the vacuum insulated cryostat, made from the low activity stainless steel of type 1.4571/316Ti (316Ti SS). The target consists of 62~kg of LXe, defined by a structure made from polytetrafluoroethylene (PTFE, {\it Teflon}) and copper.
The target volume is viewed by two arrays of photomultiplier tubes (PMT), one on the bottom immersed in LXe, and one in the gas phase above the target volume.
The electric fields in the TPC are generated by applying potential differences across the electrodes, which are made of stainless steel meshes welded onto 316Ti SS rings. They include two electrodes on the bottom of the TPC above the bottom PMT array, and a stack of three electrodes at the liquid-gas interface.

\begin{figure}[!b]
\resizebox{\columnwidth}{!}{\includegraphics{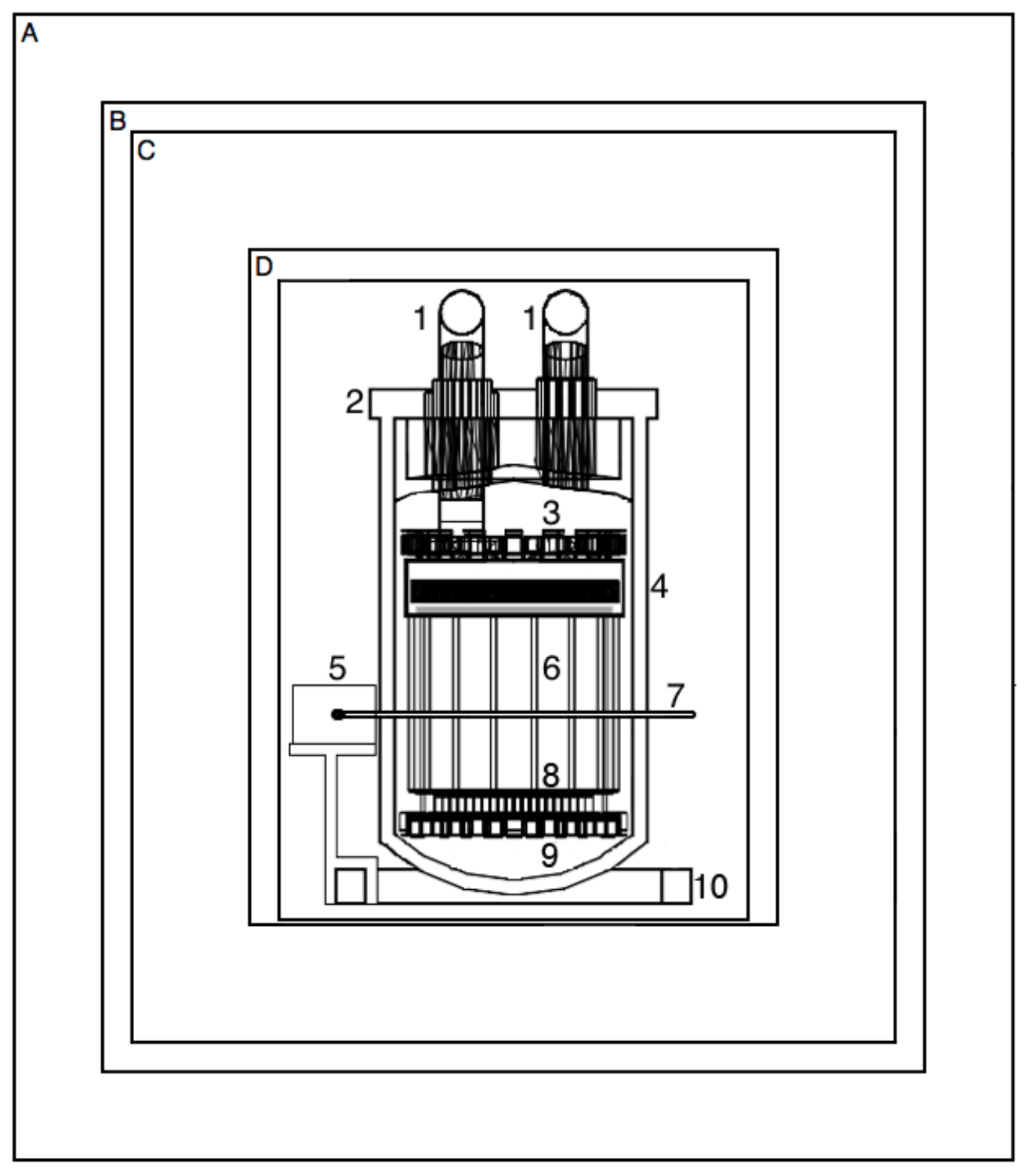}}
\caption{The GEANT4 model of the XENON100 detector and its shield: A - outer lead layer, B - inner lead layer with low $^{210}$Pb contamination, C - polyethylene shield, D - copper shield; 1 -  pipes to the PMT feedthroughs and pumping ports, 2 - stainless steel cryostat, 3 - top and side top PMT arrays in the veto, 4 - top PMT array in the TPC, in the gas phase inside the 'diving bell', 5 - lead brick for calibration with $^{241}$AmBe neutron source, 6 - TPC wall (PTFE panels), 7 - copper pipe for calibration sources, 8 - bottom PMT array in the TPC, 9 - bottom and side bottom PMT array in in the veto, 10 - support bars for the cryostat. 
The water shield and an additional polyethylene layer on the bottom are not shown.}
\label{DetectorModel}
\end{figure}

In order to simulate the response of the detector to various types of particles and to predict the intrinsic and ambient electronic recoil background, a detailed model (Figure~\ref{DetectorModel}) has been created with the GEANT4 toolkit~\cite{g4}. Table~\ref{tab:ScreeningResults} shows the amount of materials used for the detector construction, computed from the model and in agreement with the actual detector.

\begin{table*}[!ht]
{\centering
\caption{Materials used to construct the XENON100 detector and shield, and their radioactive contamination from measurements at underground facilities at LNGS~\cite{screening}. The cryostat vessels with the top flange and pipes, and the 'diving bell' system are made from the grade 316Ti SS and shown as one unit. The resistive voltage divider network for the TPC drift field is simplified in the model with a thin tube. The PMT bases made from $Cirlex$ have been screened fully assembled, with the resistors and capacitors.}
\label{tab:ScreeningResults}
\vspace{0.2cm}
\begin{tabular}{l|l|c|c|c|c|l}
\hline
Component 					& Amount 		& \multicolumn{5}{c}{Total radioactive contamination in materials [mBq/amount]} \\
							&			& $^{238}$U / $^{226}$Ra	& $^{232}$Th 	 	& $^{60}$Co 	 	& $^{40}$K 	 	& other nuclides \\
\hline
Cryostat and 'diving bell' (316Ti SS) 	& 73.61 kg 	& $<$ 130			& $<$ 140			& 400$\pm$40		&  $<$ 660 		&\\
Support bars (316Ti SS) 			& 49.68 kg 	& $<$ 65 			& 140$\pm$30		& 700$\pm$20 		& $<$ 350 		&\\
Detector PTFE 					& 11.86 kg	& $<$ 0.7 			& $<$ 1.2			&  $<$ 0.4			&  $<$ 8.9 		&\\
Detector copper 				& 3.88 kg 		& $<$ 0.86 		& $<$ 0.62 		& 0.78$\pm$0.31 	& $<$ 5.20 		&\\
PMTs						& 242 pieces 	& 36$\pm$5		& 41$\pm$10		& 150$\pm$20 		& 2700$\pm$500 	& $^{137}$Cs: $<$~190 \\
PMT bases 					& 242 pieces 	& 39.0$\pm$5.0	& 17.0$\pm$5.0	& $<$2.4			& $<$39.0 		&\\
TPC resistor chain 				& 1.5$\times$10$^{-3}$ kg 		& 1.10$\pm$0.20 	& 0.57$\pm$0.12	& $<$ 0.12		& 7.8$\pm$1.2 		&\\
Bottom electrodes (316Ti SS) 		& 0.23 kg 		& 0.81$\pm$0.06	& 0.39$\pm$0.04 	& 1.60$\pm$1.00	& $<$ 1.10 		&\\
Top electrodes (316Ti SS) 		& 0.24 kg 		& $<$ 0.64		& $<$ 0.35 		& 3.10$\pm$0.20 	& $<$ 2.80 		&\\
PMT cables					& 1.80 kg 		& $<$ 2.9 			& 6.7$\pm$3.2		& $<$ 1.2 			& 63.0$\pm$23.0 	& $^{108m}$Ag: 9.0$\pm$1.6 \\
\hline
Copper shield 					& 2.1$\times$10$^{3}$~kg 	& 170$\pm$50 			& 25$\pm$11		& 82$\pm$12 		& 7$\pm$2		&\\ 
Polyethylene shield 				& 1.6$\times$10$^{3}$ kg 	& 370$\pm$80 			& $<$ 150 		& - 				& 1100$\pm$600 	&\\
Lead shield  (inner layer) 			& 6.6$\times$10$^{3}$ kg 	&  $<$ 4400 			&  $<$ 3600		& $<$ 730 		& $<$ 9600 		& $^{210}$Pb: (1.7$\pm$0.4)$\times$10$^{8}$ \\
Lead shield (outer layer) 			& 27.2$\times$10$^{3}$ kg 	&  $<$ 25000			& $<$ 19600		&  $<$ 3300		& 380$\pm$80		& $^{210}$Pb: (1.4$\pm$0.2)$\times$10$^{10}$ \\
\hline
\end{tabular}
}
\end{table*}

\begin{table}[!ht]
\centering
\caption{Mass model of the R8520-06-AL PMT \cite{PMTmassModel}. The last two materials are not included in the GEANT4 model due to their low mass. }
\label{tab:PMTmassModel}
\vspace{0.2cm}
\begin{tabular}{l|l|c}
\hline
PMT part 					& Material				& Weight [g] \\
\hline
Metal package and stem pins	& Kovar metal 			& 13.0	\\
Electrodes				& stainless steel 		& 7.0 \\
Glass for window			& synthetic silica		& 2.0 \\
Glass in stem				& borosilicate glass		& 1.0 \\
Aluminum ring				& Al 					& 0.1	 \\
Insulator					& ceramic				& 0.04 \\
Getter					& ZrAl				& 0.02 \\
\hline
\end{tabular}
\end{table}

The passive shield, with 4$\pi$ coverage of the detector, is installed on a 25~cm polyethylene slab (not shown). From outside to inside, it consists of tanks filled with water (thickness 20~cm, not shown) to shield against ambient neutrons, placed on 4 sides of the shield box. After the water shield, there are two layers of lead: a 15~cm outer layer and a 5~cm inner layer with low level of the radioactive isotope ${^{210}}$Pb (Table \ref{tab:ScreeningResults}). Inside the lead box, there are 20~cm of polyethylene against neutron background. The innermost shield layer consists of 5~cm thick (0.5~cm on the bottom) copper plates. It reduces the gamma background from the outer shield layers. Highly radioactive detector components are mounted outside of the shield, for example signal and high voltage feedthroughs, vacuum pumps, pressure sensors and associated electronics. An innovative detector design feature, which has contributed to the low background rate of XENON100, is the mounting of the cryogenics system, based on a pulse tube refrigerator, outside the passive shield, far from the LXe target.

The cylindrical TPC is formed by 24 interlocking PTFE panels. PTFE reflects scintillation light with high efficiency for vacuum ultraviolet~\cite{yamashita}, and optically separates the 62~kg target volume from the surrounding LXe, which is in average 4~cm thick and has a total mass of 99~kg. This allows to exploit the self-shielding capability of LXe due to its high density and high atomic number. In addition, this LXe volume around the target is instrumented with PMTs, becoming an active veto for background reduction by rejecting events in which a particle deposits part of its energy in the veto volume. 

The cryostat is supported inside the shield by the 316Ti SS bars, which are mounted onto the movable shield door. The thickness of the inner and outer cryostat walls is 1.5~mm and the total weight of the vessel is 70.0~kg, which is only 30\% of that of the XENON10 detector's cryostat~\cite{xe10-instrument}. The inner vessel containing the LXe is lined on the walls and the bottom with a 1.5~mm thick PTFE layer  in order to increase the light collection efficiency in the active veto volume.

Electrons created by ionization in the LXe target are drifted upwards by a strong electric field created by applying voltage on the cathode, installed on the bottom of the TPC. In order to shield the bottom PMTs from this electric field, an additional grounded electrode is installed below the cathode.

The gas phase for charge amplification is maintained using a 'diving bell' system, made from 316Ti SS with a total weight 3.6~kg. It allows to keep the liquid level constant at a precise height while having an additional layer of LXe above the TPC. A slight overpressure in the bell is provided by the gas returning from the continuous recirculation system, and the height of the gas outlet from the bell can be changed by a motion feedthrough in order to adjust the liquid level.

An extraction field is created across the liquid-gas interface by applying high voltage on the anode, which is placed inside the 'diving bell'. 
Two additional electrodes are installed below and above the anode and are kept at ground potential in order to close the field cage and shield the top PMT array from the high electric field. Ionization electrons are extracted into the gas phase and accelerated, producing the proportional scintillation (S2) signal \cite{S2}. The gaps between the top electrodes are 5~mm, and the liquid level is adjusted between the lower two of them. In the background model, only the 316Ti SS support rings for the electrodes are considered, given that the meshes are $\sim$100~$\mu$m thick and have a very low mass, leading to a negligible background from their radioactivity.

The prompt (S1) and the proportional scintillation light (S2) is detected by 242 1"-square R8520-06-AL Hamamatsu PMTs. They are among the lowest radioactivity PMTs, and are optimized to operate in LXe. The top PMT array consists of 98 PMTs, mounted in a concentric pattern in a PTFE support structure inside the 'diving bell'. The bottom PMT array consists of 80 PMTs, mounted below the cathode in the LXe, and arranged in a rectangular pattern in order to maximize the photocathode coverage. Additionally, 64 PMTs view LXe of the veto volume: 16 PMTs above and below the TPC, and 32 observing the sides.
The components of the PMT are shown in Table~\ref{tab:PMTmassModel}.
In the GEANT4 model, a PMT is simplified with a stainless steel case and a synthetic silica window inside a thin aluminum ring.
A PMT is supplied with high voltage through the voltage divider circuit mounted on a base made from $Cirlex$. The base is approximated in the model as a homogenous unit.

The XENON100 data acquisition system (DAQ) digitizes the full waveform of the 242 PMTs at 100~MHz, where the time window for an event is 400~$\mu$s, more than twice the maximum electron drift time. If a particle has deposited energy at multiple places in the target, then two or more S2 pulses are recorded in the trace. Such an event is a multiple scatter event and is rejected in the analysis since the predicted behavior of the WIMP, due to its very low scattering cross-section, is a single scatter event.

For the calculation of the final background rate in the Monte Carlo simulations, multiple scatter events are rejected taking into account the finite position resolution of the detector.  
A multiple scatter event is considered as a single scatter event if the interactions happen less than 3~mm apart in $Z$. This position resolution is given by the width of the S2 signals and the peak separation efficiency of the S2 peak finder algorithm.

\section{Background due to radioactive contamination in the detector and shield materials}
\label{secDetectorMaterials}

Special care has been taken to select detector and shield materials according to their radioactive contamination. Before detector construction, the majority of materials planned to be used were screened with low background Ge detectors in order to determine their intrinsic radioactivity, mostly due to residual $^{238}$U, $^{232}$Th, $^{40}$K, and $^{60}$Co contamination.
XENON has access to a dedicated screening facility underground at LNGS, the Gator detector~\cite{gator}. Moreover, the LNGS screening facility, with some
of the most sensitive Ge detectors in the world \cite{LNGSfacility}, has also been used.
The radioactive contamination of the materials used for detector and shield construction is shown in Table~\ref{tab:ScreeningResults}. For more details, see Ref.~\cite{screening}.

Decays of the radioactive isotopes in the materials listed in Table~\ref{tab:ScreeningResults} have been simulated with GEANT4, and the corresponding background rates have been calculated. The measured activities have been used as an input information for the Monte Carlo simulations and background predictions. For the analysis presented here, the upper limits are treated as detection values.

\begin{figure}[!t]
\includegraphics[height=0.73\linewidth]{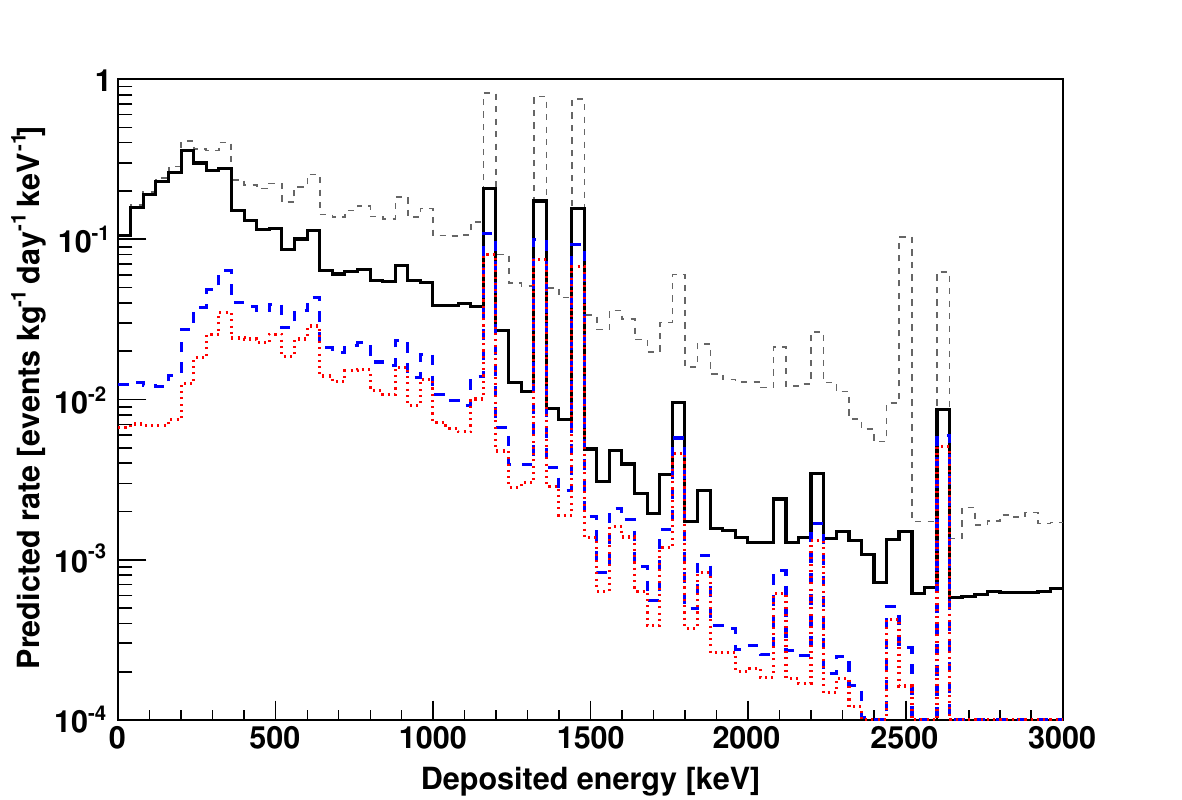}
\caption{(Color online) Predicted background from the detector and shield materials: energy spectra of all events (thin dashed line) and single scatters (solid line) in the entire 62~kg LXe target, and single scatters in the 40~kg (thick dashed line) and 30~kg fiducial volumes (dotted line), with infinite energy resolution.}
\label{figSpectraFS}
\end{figure}

\begin{figure}[!t]
\includegraphics[height=0.73\linewidth]{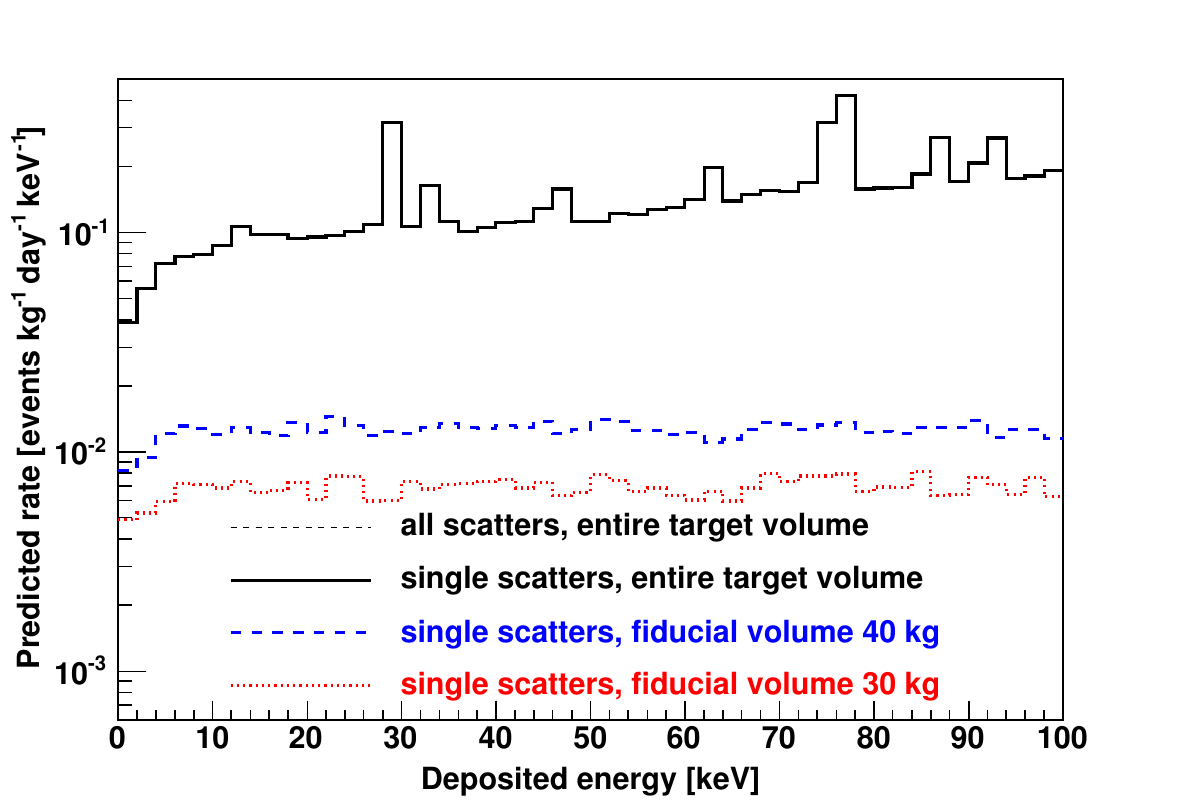}
\caption{Zoom into the low energy region of the Monte Carlo spectra shown in Figure~\ref{figSpectraFS}. The spectra of all scatters and single scatter events in the entire target volume overlap.}
\label{figSpectraWS}
\end{figure}

Figure~\ref{figSpectraFS} shows the spectra in the entire energy range, and Figure~\ref{figSpectraWS} in the region of interest. The energy range for the background rate calculation is chosen to be sufficiently wide, up to 100~keV, to include the signal region for inelastic dark matter which is predicted to be in a higher energy range than the one from standard elastic WIMP scattering~\cite{inelastic}.  The effect of the discrimination between multiple and single scatter events on the background rate can be seen in Figure~\ref{figSpectraFS}: the multiple scatter behavior of incident gamma rays is typical for higher energies, whereas single scatter events dominate in the low energy region, and the multiple scatter cut does not yield a significant reduction of the background rate. Further background reduction can be achieved with fiducial volume cuts.

In Figure~\ref{figSpectraWS} several characteristic X-rays can be seen. The xenon K-shell fluorescence peaks appear at 30~keV and 34~keV. The X-ray peaks at 15, 75, 85 and 90~keV are from Pb and Bi close to the target volume, for example in PTFE walls. In addition, there is a 46~keV gamma line from $^{210}$Pb decay, and 63~keV gamma line from the decay of $^{234}$Th. Due to their short mean free path, these low energy lines can be observed only at the edge of the LXe volume. After applying a cut on the position of the interactions, the peaks disappear and the spectrum becomes relatively flat in the low energy region. The background rate is thus presented as the average below 100~keV.

\begin{figure}[!h]
\includegraphics[height=0.73\linewidth]{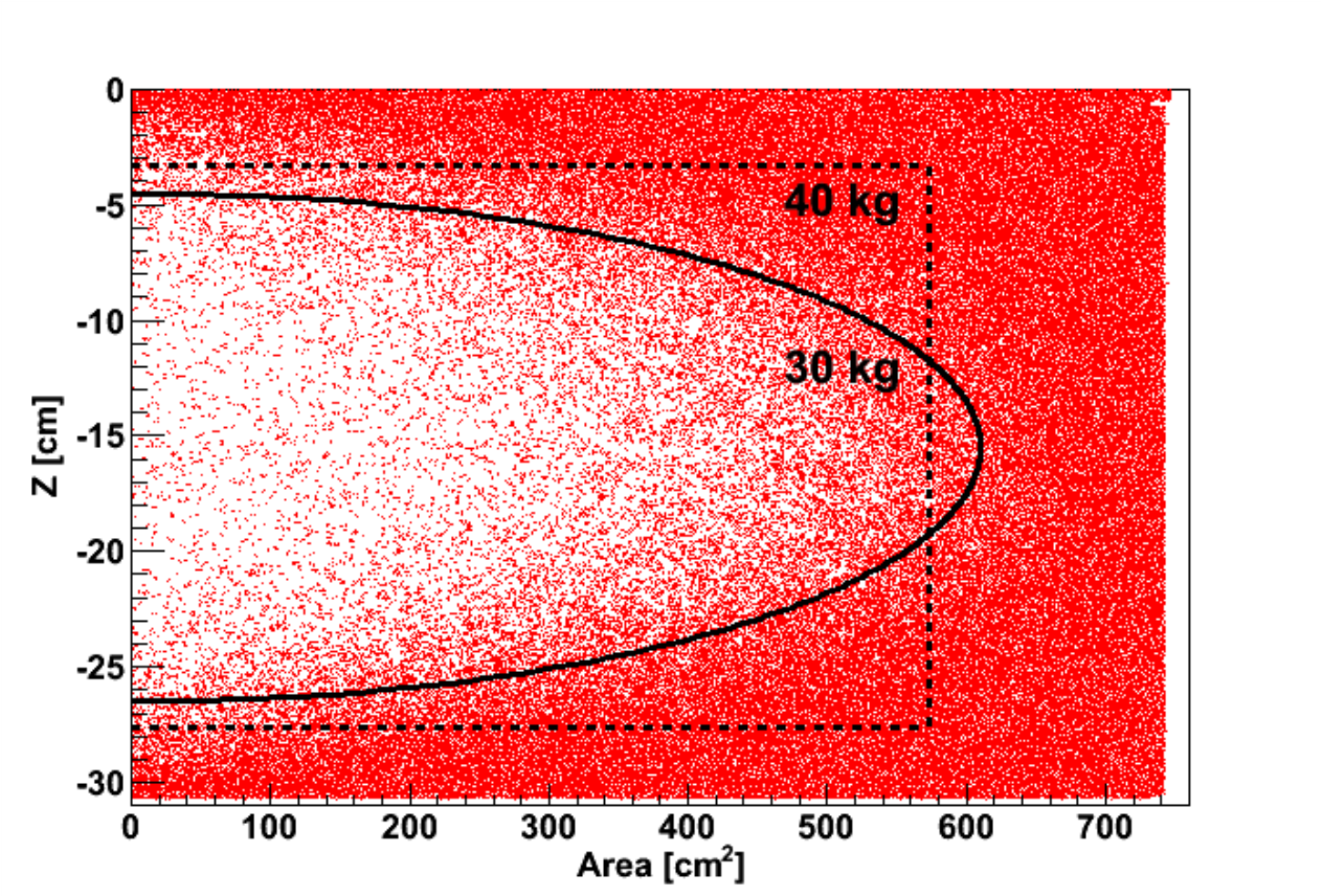}
\caption{(Color online) Predicted electronic recoil background from detector and shield materials, excluding intrinsic radioactivity in the LXe. Shown are single scatter events with energy below 100~keV in the TPC, without veto cut. $Z$ = 0~cm corresponds to the liquid-gas interface. The cathode mesh is located at $Z$ = $-$304.5~mm. The dashed line shows the 40~kg fiducial volume, and the solid line illustrates the 30~kg fiducial volume optimized to minimize the background.}
\label{figPositionDistribution}
\end{figure}

The spatial distribution of the single scatter electronic recoil events in the region of interest is presented in Figure~\ref{figPositionDistribution}. The radial cut rejects events at the edge of the target volume, originating mostly from radioactive decays in the PTFE of the TPC and the 316Ti SS of the cryostat vessels.
The background from the PMTs, PMT bases, 'diving bell', and the electrodes can be efficiently reduced by rejecting events within the top and bottom layers of LXe.

\begin{table*}[!ht]
\centering
\caption{Predicted rate of single scatter electronic recoil events in the energy region below 100~keV, in the entire target volume and in 40~kg and 30~kg fiducial volumes, with and without the veto cut with an average energy threshold of 100~keV. The statistical errors of the simulation are less than 1\%. The background from the support rings for the anode stack meshes is relatively high only within a few mm near the liquid-gas interface, hence it can be sufficiently reduced with the fiducial volume cuts. The background from the lead shield is negligible.}
\label{tab:gamma-rates}
\vspace{0.2cm}
\begin{tabular}{l|cc|cc|cc}
\hline
& \multicolumn{6}{c}{Single electronic recoils [$\times$10$^{-3}$~events$\cdot$kg$^{-1}$$\cdot$day$^{-1}$$\cdot$keV$^{-1}$]}\\
\hline
Volume  & \multicolumn{2}{c|}{62~kg target} & \multicolumn{2}{c|}{40~kg fiducial}   & \multicolumn{2}{c}{30~kg fiducial} \\
\hline
Veto cut       & \multicolumn{1}{c|}{none} & \multicolumn{1}{c|}{active} & \multicolumn{1}{c|}{none} & \multicolumn{1}{c|}{active}  & \multicolumn{1}{c|}{none} & \multicolumn{1}{c}{active} \\
\hline
Cryostat and 'diving bell' (316Ti SS) 	& 20.85	& 6.70	& 2.62	& 0.65 	& 1.81	& 0.48 \\
Support bars (316Ti SS)			& 1.05 	& 0.24	& 0.19	& 0.05 	& 0.13	& 0.04 \\
Detector PTFE 					& 3.47 	& 2.89	& 0.05 	& 1.5$\times$10$^{-2}$ 	& 3.4$\times$10$^{-2}$	& 1.1$\times$10$^{-3}$ \\
Detector copper 				& 0.31  	& 0.13	& 0.02	& 4.7$\times$10$^{-3}$ & 1.2$\times$10$^{-2}$ & 2.6$\times$10$^{-3}$ \\
PMTs 						& 81.57	& 44.48	& 8.84	& 2.40 	& 4.61	& 1.32 \\
PMT bases	 				& 15.95	& 10.26	& 0.86	& 0.22 	& 0.40 	& 0.12 \\
TPC resistor chain 				& 1.7$\times$10$^{-4}$ & 1.2$\times$10$^{-4}$  &  2.7$\times$10$^{-6}$ &  7.1$\times$10$^{-7}$ & 2.1$\times$10$^{-6}$ 	& 5.7$\times$10$^{-7}$ \\
Bottom electrodes (316Ti SS) 		& 0.73 	& 0.35	& 0.03	& 6.4$\times$10$^{-3}$ 	& 0.02 	& 4.1$\times$10$^{-3}$ \\
Top electrodes (316Ti SS) 		& 11.76 	& 9.86	& 0.03	& 7.0$\times$10$^{-3}$ 	& 0.02 	& 4.6$\times$10$^{-3}$ \\
PMT cables 					& 0.56	& 0.08	& 0.10	& 0.02 	& 0.08 	& 0.02 \\
Copper shield					& 0.64   	& 0.22	& 0.10	& 0.04 	& 0.07 	& 0.02 \\
Polyethylene shield				& 0.33   	& 0.09	& 0.05	& 0.01 	& 0.03 	& 0.01 \\
\hline
Total 						& 137.22  	& 75.30  	&12.89  	& 3.42 	& 7.22 	& 2.02 \\
\hline
\end{tabular}
\end{table*}

\begin{figure}[!t]
\includegraphics[height=0.73\linewidth]{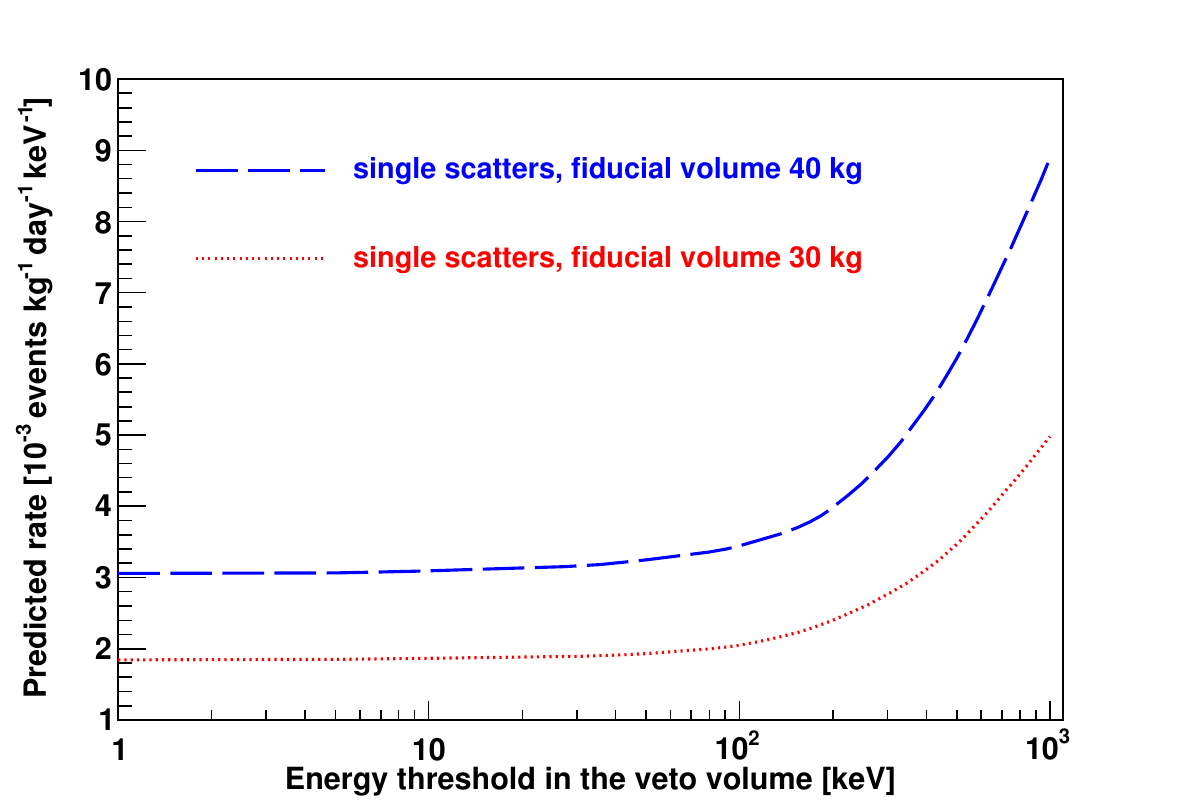}
\caption{(Color online) Predicted background rate from the detector and shield materials in the energy range below 100~keV, as a function of the energy threshold in the active veto. The average energy threshold in the veto measured with a collimated $^{137}$Cs source is about 100~keV.}
\label{figVetoCuts}
\end{figure}

The effect of the active LXe veto is presented in Figure~\ref{figVetoCuts}, showing the total rate as a function of the energy threshold in the veto volume.
The measured efficiency of the veto coincidence cut has been implemented in the Monte Carlo simulations. The average energy threshold measured with a collimated $^{137}$Cs source is about 100~keV. This allows to reduce the background rate in the entire target volume by $\sim$50\%. Background reduction is even more efficient if the veto cut is combined with a fiducial volume cut, which results in a $>$90\% reduction of the background rate. The reduction of the background rate remains almost constant when the energy threshold in the veto is below 100~keV. This is explained by an anti-correlation of the energy deposition in the active veto and target volume: events that deposit a small amount of energy in the target volume are likely to have deposited a larger amount of energy in the veto volume.

Table~\ref{tab:gamma-rates} presents the energy averaged rates of single scatter electronic recoil from detector materials in the region of interest, below 100~keV. The background rate has been predicted for the entire 62~kg LXe target, and for two fiducial volumes: a simple 40~kg cylindrical fiducial volume used in the analysis of the first XENON100 data \cite{xe100-independent}, and a 30~kg fiducial volume cut optimized to minimize the background.

\begin{figure}[!t]
\includegraphics[height=0.73\linewidth]{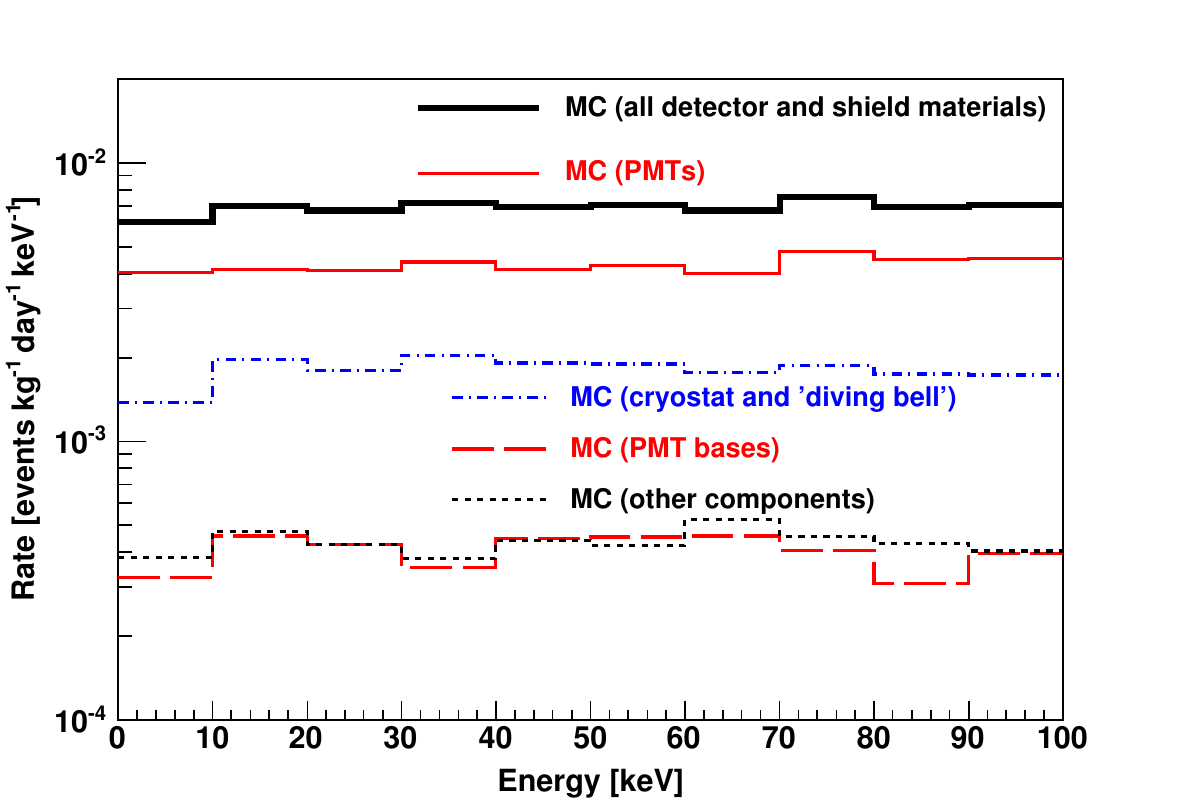}
\caption{(Color online) The predicted background from the detector and shield materials (thick black line) in the 30~kg fiducial mass without veto cut, together with the individual contributions from the PMTs (solid line), the cryostat with pipes and 'diving bell' (dash dotted line), PMT bases (long dashed line). The short dashed line shows the summed background from all other components: detector PTFE and copper, cryostat support bars, TPC resistor chain, top and bottom electrodes, PMT cables, and copper and polyethylene shield.}
\label{figSpectraDetectorMaterials}
\end{figure}

The low energy Monte Carlo spectrum of the background from the detector and shield materials is shown together with the individual contributions in Figure~\ref{figSpectraDetectorMaterials}, for the 30~kg fiducial mass without veto cut. The background rate is dominated by the PMTs ($\sim$65\% of the total background from all detector and shield materials), and the 316Ti SS cryostat, pipes and 'diving bell' (other $\sim$25\%). The dominant contribution to the background from the PMTs is originating from the $^{60}$Co and $^{40}$K contamination (50\% and 34\%, respectively). The main contaminant in the 316Ti SS is  $^{60}$Co, which is responsible for almost 70\% of the total background from this material. Components as detector PTFE and copper, cryostat support bars, TPC resistor chain, top and bottom electrodes, PMT cables, copper and polyethylene shield contribute $<$10\% to the total background rate from the detector and shield materials.

\section{Background from radon in the shield cavity}

\label{secRadonCavity}

\begin{figure}[!b]
\resizebox{\columnwidth}{!}{\includegraphics[height=0.82\linewidth]{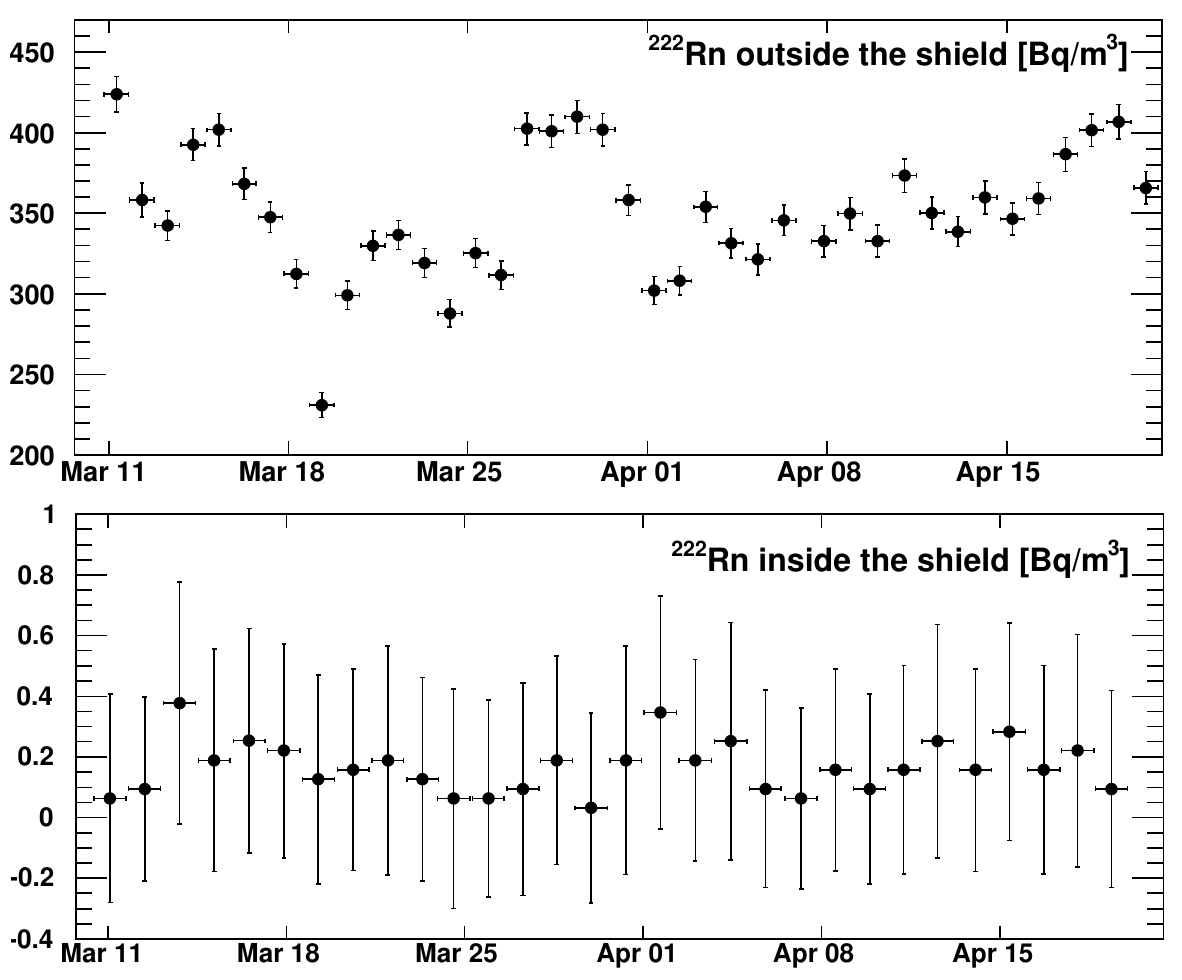}}
\caption{(Color online) $^{222}$Rn activity measured during 6 weeks of a science run in 2010, at the site of the experiment (top) and inside the shield (bottom). Each datapoint shows measurements averaged over 24~hours. No correlation can be observed. The measurements inside the cavity are at the sensitivity limit of the radon monitor. }
\label{figRadonA}
\end{figure}

A potentially dangerous background for XENON100 is the gamma background from the
decay of ${^{222}}$Rn daughters inside the shield cavity. 
The average measured radon activity in the LNGS tunnel at the location of the experiment is $\sim$350~Bq/m$^3$ (Figure~\ref{figRadonA}, top). 
Therefore, the shield cavity with a total volume 0.58~m$^3$ is constantly flushed with nitrogen gas when the shield door is closed. Nevertheless, a certain
amount of radon can still be present. During the science runs, a low and constant $^{222}$Rn concentration is kept inside the shield.  It is continuously monitored as shown in Figure~\ref{figRadonA}, bottom. The measured values are at the limit of the sensitivity of the radon monitor. No correlation can be seen between the radon concentration inside and outside the shield.

\begin{figure}[!h]
\includegraphics[height=0.73\linewidth]{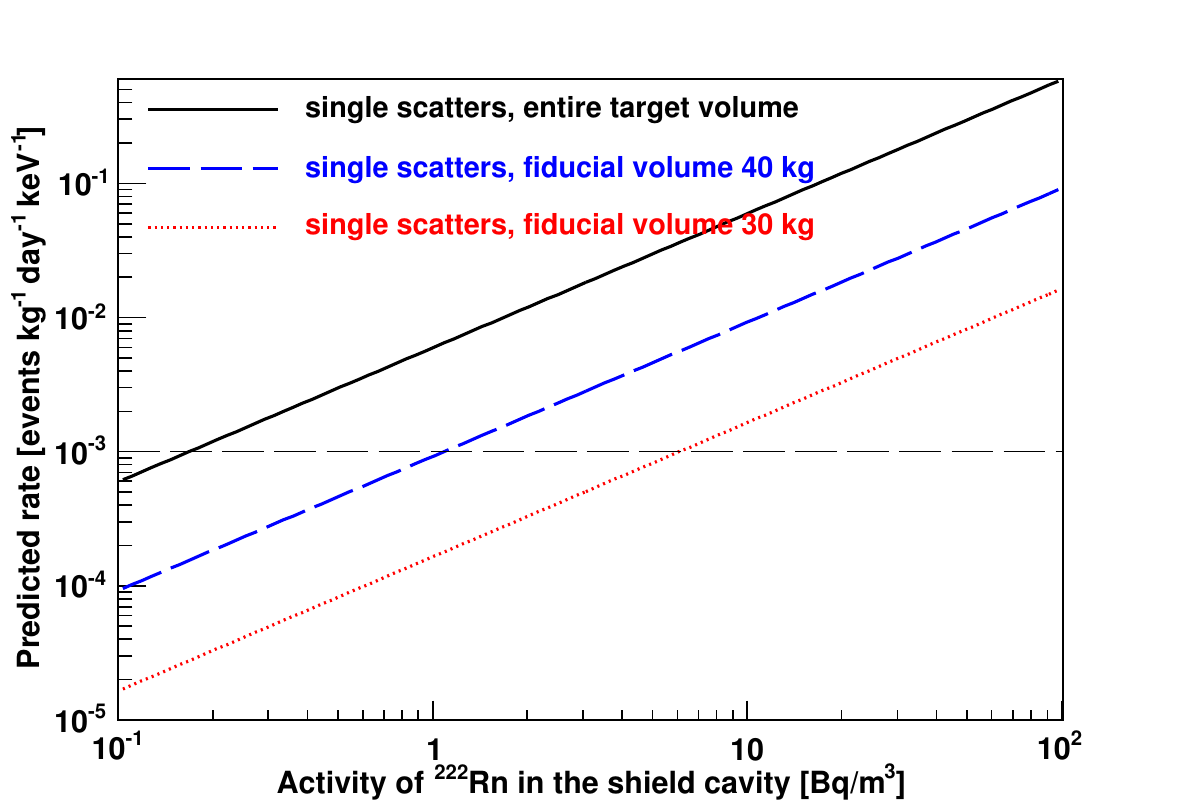}
\caption{(Color online) Predicted rate of single electronic recoils with energy below 100~keV as a function of $^{222}$Rn concentration in the shield cavity for different fiducial masses. As a reference value, the horizontal dashed line corresponds to a background rate of 10$^{-3}$~events$\cdot$kg$^{-1}\cdot$day$^{-1}\cdot$keV$^{-1}$.}
\label{figRadonB}
\end{figure}

Figure~\ref{figRadonB} shows the predicted background rate due to $^{222}$Rn as a function of its concentration inside the shield. Without veto cut, the background rate from 1~Bq/m$^{3}$ of $^{222}$Rn in the shield is 6$\times$10$^{-3}$~events$\cdot$kg$^{-1}\cdot$day$^{-1}\cdot$keV$^{-1}$ for the entire target mass of 62~kg, 9$\times$10$^{-4}$~events$\cdot$kg$^{-1}\cdot$day$^{-1}\cdot$keV$^{-1}$ in the 40~kg fiducial volume, and 2$\times$10$^{-4}$~events$\cdot$kg$^{-1}\cdot$day$^{-1}\cdot$keV$^{-1}$ in the 30~kg fiducial volume. For the 30~kg fiducial mass, this is less than 2\% of the background from the detector and shield materials. Moreover, the measured radon concentration is well below 1~Bq/m$^{3}$.

\section{Background due to intrinsic krypton and radon radioactivity}
\label{secIntrinsicBG}

There is no long-lived radioactive xenon isotope, with the exception of the potential double beta emitter $^{136}$Xe, with the half-life limits of $>$7$\times$10$^{23}$~years and $>$1.1$\times$10$^{22}$~years for the neutrinoless and 2$\nu$ double beta decay, respectively~\cite{DoubleBetaLimit}.

Commercially available xenon gas, where purification is performed by distillation and adsorption-based chromatography, has a concentration of natural krypton at the ppm level.
Natural krypton contains about 10$^{-11}$~g/g of radioactive $^{85}$Kr. 
The background from the beta decay of $^{85}$Kr, with T$_{1/2}$ = 10.76~years and endpoint energy 687~keV, is a potential limitation in the sensitivity of rare-event searches using xenon targets.

The gas used in the XENON100 experiment has been processed at a commercial distillation plant to reduce the concentration of krypton to $<$10~ppb.
The high-temperature getter used in the experiment to purify xenon from water and electronegative contaminants does not remove the noble gas krypton. Therefore, an additional gas purification has been performed with cryogenic distillation. The reduction of krypton concentration down to a few ppt has been reported in Ref.~\cite{distillation}, with a distillation column similar to the one procured by XENON100.

\begin{figure}[!h]
\includegraphics[height=0.73\linewidth]{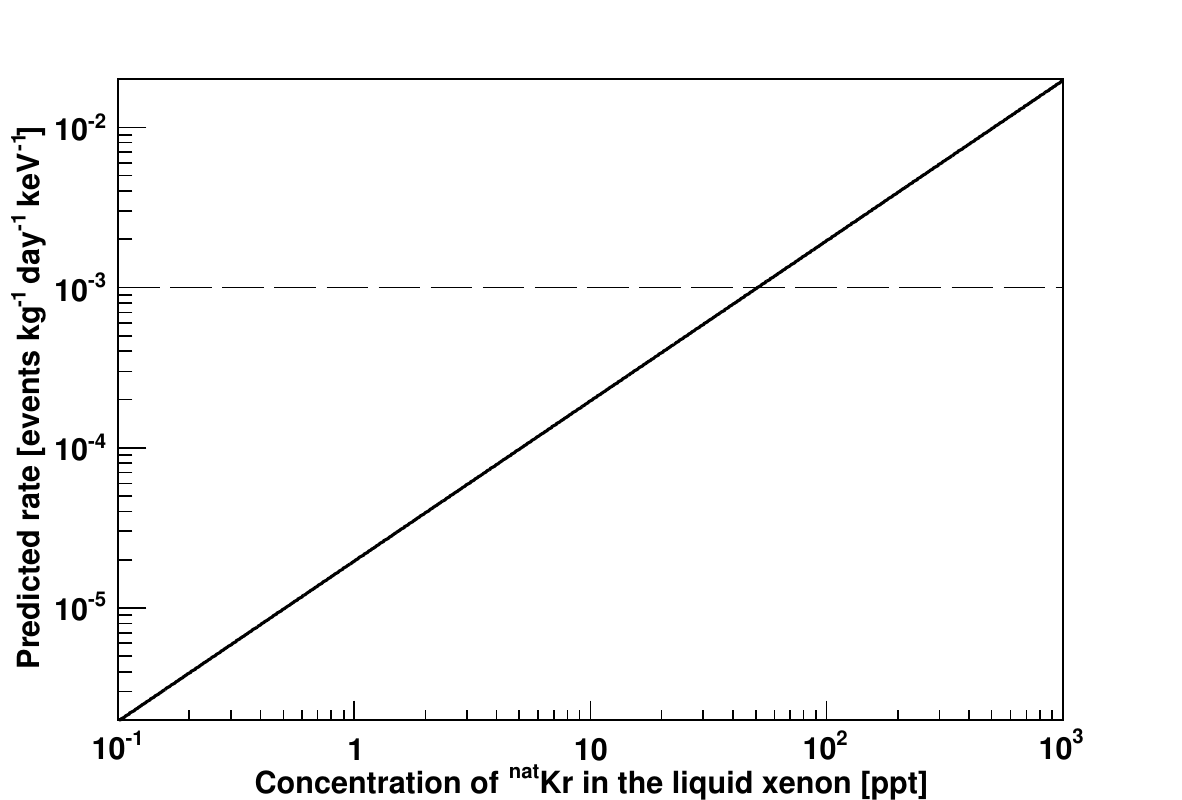}
\caption{Rate of single electronic recoils from $^{85}$Kr decay in the energy region below 100~keV as a function of the concentration of natural Kr in the LXe. Fiducial and veto cuts are inefficient for this intrinsic background source. As a reference value, the horizontal dashed line corresponds to a background rate of 10$^{-3}$~events$\cdot$kg$^{-1}\cdot$day$^{-1}\cdot$keV$^{-1}$. }
\label{figLXeKr}
\end{figure}

Levels of radioactive trace contaminations in xenon might vary at different stages of the experiment, as they strongly depend on purification processes. The background rate from intrinsic radioactivity in LXe has thus been predicted for different  concentrations of $^{nat}$Kr in LXe, and is shown in Figure~\ref{figLXeKr}.

Another intrinsic source of background is the decay of $^{222}$Rn daughters in the LXe. Radon is present in the LXe due to emanation from detector materials and the getter, and diffusion of the gas  through the seals. 

\begin{figure}[!b]
\includegraphics[height=0.73\linewidth]{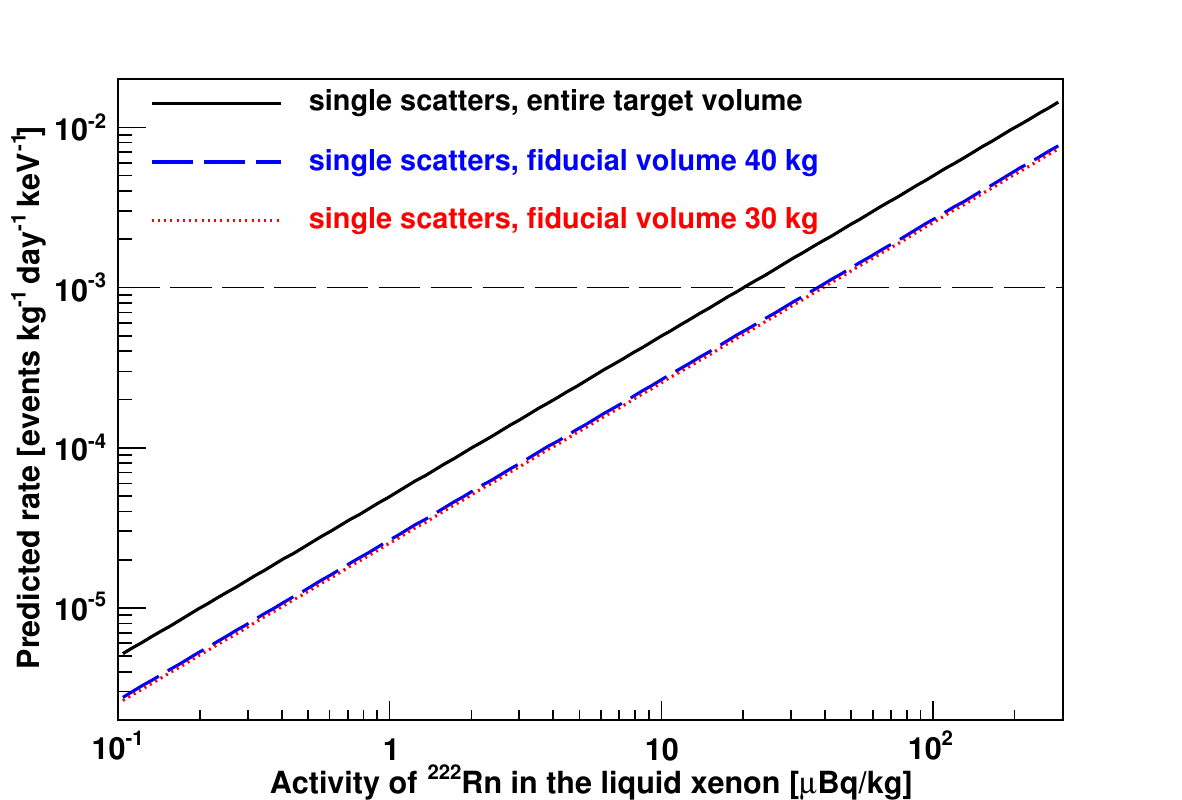}
\caption{(Color online) Predicted background rate below 100~keV as a function of $^{222}$Rn concentration in the LXe. As a reference value, the horizontal dashed line corresponds to a background rate of 10$^{-3}$~events$\cdot$kg$^{-1}\cdot$day$^{-1}\cdot$keV$^{-1}$. }
\label{figLXeRn}
\end{figure}

In the Monte Carlo simulation, $^{222}$Rn decays are generated uniformly in the LXe, and only the part of the chain before $^{210}$Pb is considered, since the relatively long half-life time of 22.3 years for $^{210}$Pb results in radioactive disequilibrium in the decay chain.
The predicted background rate in the energy region below 100~keV is shown in Figure~\ref{figLXeRn} as a function of the $^{222}$Rn concentration in the LXe.

\begin{figure*}[!ht]
\begin{center}
\includegraphics[width= 0.95\linewidth]{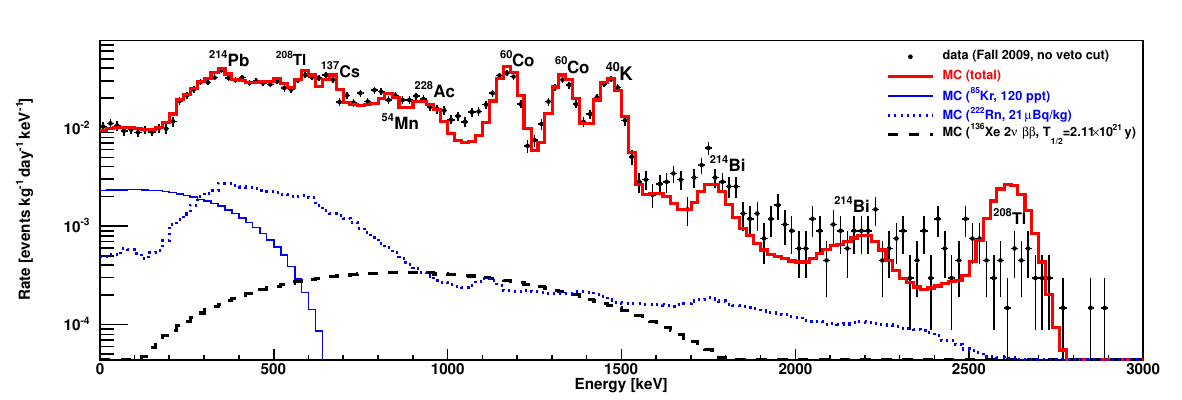}
\caption{(Color online) Energy spectra of the background from measured data (commissioning run in Fall 2009~\cite{xe100-independent}) and Monte Carlo simulations in the 30~kg fiducial volume without veto cut (thick red solid line). Cosmogenic activation of LXe is not included. The energy spectra of $^{85}$Kr and  $^{222}$Rn decays in LXe are shown with the thin blue solid and dotted lines, respectively. The thin black dashed histogram shows the theoretical spectrum of the 2$\nu$ double beta decay of $^{136}$Xe, assuming half-life of 2.11$\times$10$^{21}$~years~\cite{EXO}. }
\label{figDataMC}
\end{center}
\end{figure*}


A background contribution from each intrinsic radioactive source of less than 10$^{-3}$~events$\cdot$kg$^{-1}\cdot$day$^{-1}\cdot$keV$^{-1}$, which is used as a reference value, translates into a concentration of $^{nat}$Kr below 50~ppt, and an activity of $^{222}$Rn in LXe of $<$20~$\mu$Bq/kg in the entire target mass of 62~kg. The background from $^{222}$Rn daughters in the LXe can be reduced by a fiducial volume cut, removing decays at the edge of the target volume which are likely to produce high energy gamma rays with a longer mean free path which escape the target volume. For the 40~kg and 30~kg fiducial volumes, a background level of 10$^{-3}$ events$\cdot$kg$^{-1}\cdot$day$^{-1}\cdot$keV$^{-1}$ corresponds to 35 $\mu$Bq/kg.

\section{Comparison of the predictions with the measured data}
\label{secComparison}

During the commissioning run in Fall 2009~\cite{xe100-independent}, the level of krypton in the LXe has been measured with a delayed coincidence analysis using a decay channel where $^{85}$Kr undergoes a beta-decay with E$_{max}$~=~173.4~keV to $^{85m}$Rb ($\tau$~=~1.46~$\mu$s), which in turn decays to the ground state emitting a gamma-ray with an energy of 514~keV. The concentration of $^{nat}$Kr in the LXe determined with this technique is 143$_{-90}^{+130}$~ppt [mol/mol], assuming a $^{85}$Kr abundance of 10$^{-11}$.

The $^{222}$Rn level in the LXe has been determined using a beta-alpha time coincidence analysis, where events corresponding to the decays of $^{214}$Bi (T$_{1/2}$~=~19.7~min, E$_{max}$~=~3.27~MeV) and $^{214}$Po (T$_{1/2}$~=~164~$\mu$s, E$_{\alpha}$~=~7.69~MeV) are tagged. Based on this analysis, the upper limit on the $^{222}$Rn activity in LXe is $<$21~$\mu$Bq/kg.

The volumetric activity of $^{222}$Rn inside the shield cavity has been continuously monitored during this commissioning run and was always below 1~Bq/m$^{3}$.

\begin{table*}[!ht]
\centering
\caption{Summary of the predicted electronic recoil background: rate of single scatter events in the energy region below 100~keV, before S2/S1 discrimination. The veto cut with an average energy threshold of 100~keV has been applied. }
\label{tab:summaryElectronRecoils}
\vspace{0.2cm}
\begin{tabular}{l|cc|cc|cc}
\hline
&  \multicolumn{6}{c}{Predicted rate [$\times$10$^{-3}$ events$\cdot$kg$^{-1}\cdot$day$^{-1}\cdot$keV$^{-1}$]}\\
\hline
Volume &  \multicolumn{2}{c|}{62~kg target} & \multicolumn{2}{c|}{40~kg fiducial}  & \multicolumn{2}{c}{30~kg fiducial} \\
\hline
Veto cut & \multicolumn{1}{c|}{~~none~~} & \multicolumn{1}{c|}{active} & \multicolumn{1}{c|}{~~none~~} & \multicolumn{1}{c|}{active} & \multicolumn{1}{c|}{~~none~~} & \multicolumn{1}{c}{active} \\
\hline
Detector and shield materials			& 137.22 	& 75.30	& 12.89	& 3.42  	& 7.22  	& 2.02 \\
$^{222}$Rn in the shield (1~Bq/m$^{3}$)	& 5.95 	& 1.72	& 0.92	& 0.16  	& 0.16 	& 0.02 \\
$^{85}$Kr in LXe (150~ppt of $^{nat}$Kr) 	& 2.90   	& 2.90 	& 2.90 	& 2.90 	& 2.90  	& 2.90 \\
$^{222}$Rn in LXe (21~$\mu$Bq/kg)	& 1.04   	& 0.51	& 0.56 	& 0.38       	& 0.53  	& 0.37\\
\hline
All sources  						& 147.11 	& 80.43 	& 17.27 	& 6.86     & 10.81  	& 5.31 \\
\hline
\end{tabular}
\end{table*}

\begin{figure}[!h]
\begin{center}
\includegraphics[height=0.73\linewidth]{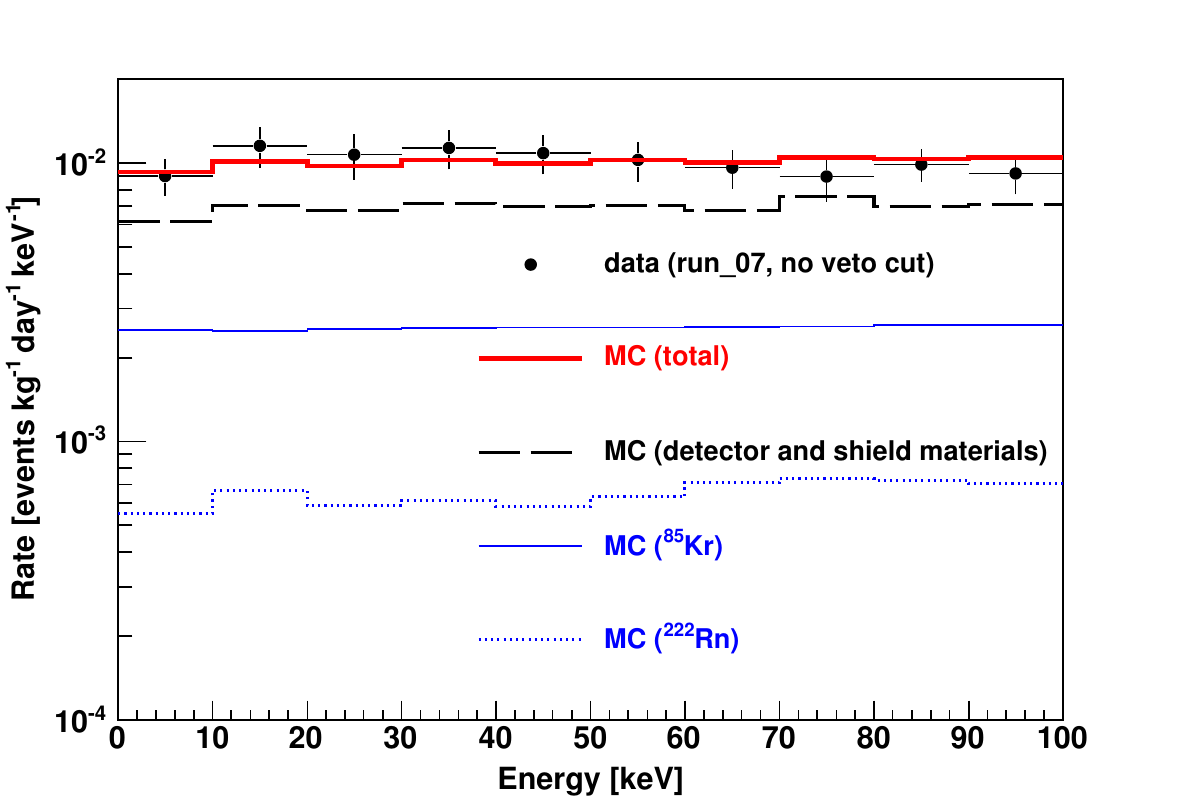}
\caption{(Color online) Zoom into the low energy region of Figure~\ref{figDataMC}: energy spectra of the measured background and Monte Carlo simulations in the 30~kg fiducial volume without veto cut. The  2$\nu~\beta\beta$ decay of $^{136}$Xe has negligible contribution to the background below 100~keV.}
\label{figDataMCzoom}
\end{center}
\end{figure}

A comparison of the measured background spectrum and the Monte Carlo simulation for the 30~kg fiducial volume without veto cut is shown in Figure~\ref{figDataMC}. The energy region below 100~keV is shown separately in Figure~\ref{figDataMCzoom}. 
For optimal energy resolution and improved linearity, the energy scale of the measured spectrum exploits the anti-correlation between the light and the charge~\cite{CombinedEnergy}: S1 and S2 are combined according to E~=~S1/4.4~+~S2/132.6~[keV]. The simulated spectrum is smeared with a Gaussian function using the energy resolution measured with calibration sources: $\frac{\sigma(E)}{E}$~=~0.009~+~0.485/$\sqrt{E~\textnormal{[keV]}}$. The contribution from the detector and shield materials is scaled based on the screening values shown in Table~\ref{tab:ScreeningResults}. The upper limits from materials screening are used as fixed values for the scaling. For the level of $^{222}$Rn in the shield cavity we used the value measured with a dedicated radon monitor. Finally, for the $^{85}$Kr and $^{222}$Rn levels in the LXe we used the values determined with the delayed coincidence analyses described above.

Very good agreement of the background model with the data is achieved for the low energy region, below 700~keV, and for the main peaks: $^{214}$Pb (352~keV), $^{208}$Tl~(583 keV), $^{137}$Cs~(662 keV), $^{60}$Co (1173 and 1332~keV), and $^{40}$K (1460~keV). In particular, simulated and measured background spectra agree well in the energy region of interest, below 100~keV (Figure~\ref{figDataMCzoom}). The predicted rates of single scatter electronic recoil events in the energy region of interest are presented in Table~\ref{tab:summaryElectronRecoils}. In the 30~kg fiducial volume, $^{85}$Kr contributes $\sim$30\% to the total background without veto cut, and 55\% when a veto coincidence cut with an average energy threshold of 100~keV is applied. The contribution from $^{222}$Rn in the LXe is $<$7\%, from $^{222}$Rn in the shield cavity $<$2\% of the total background rate in the energy region of interest.

The disagreement between simulated and measured spectra above $\sim$1.5~MeV is caused by  non-linear effects in the PMT response, which results in a worse performance of the position reconstruction algorithms, changing the rate in the fiducial volumes and leading to a worsening of the position dependent signal corrections. 

\begin{figure}[!b]
\begin{center}
\includegraphics[height=0.73\linewidth]{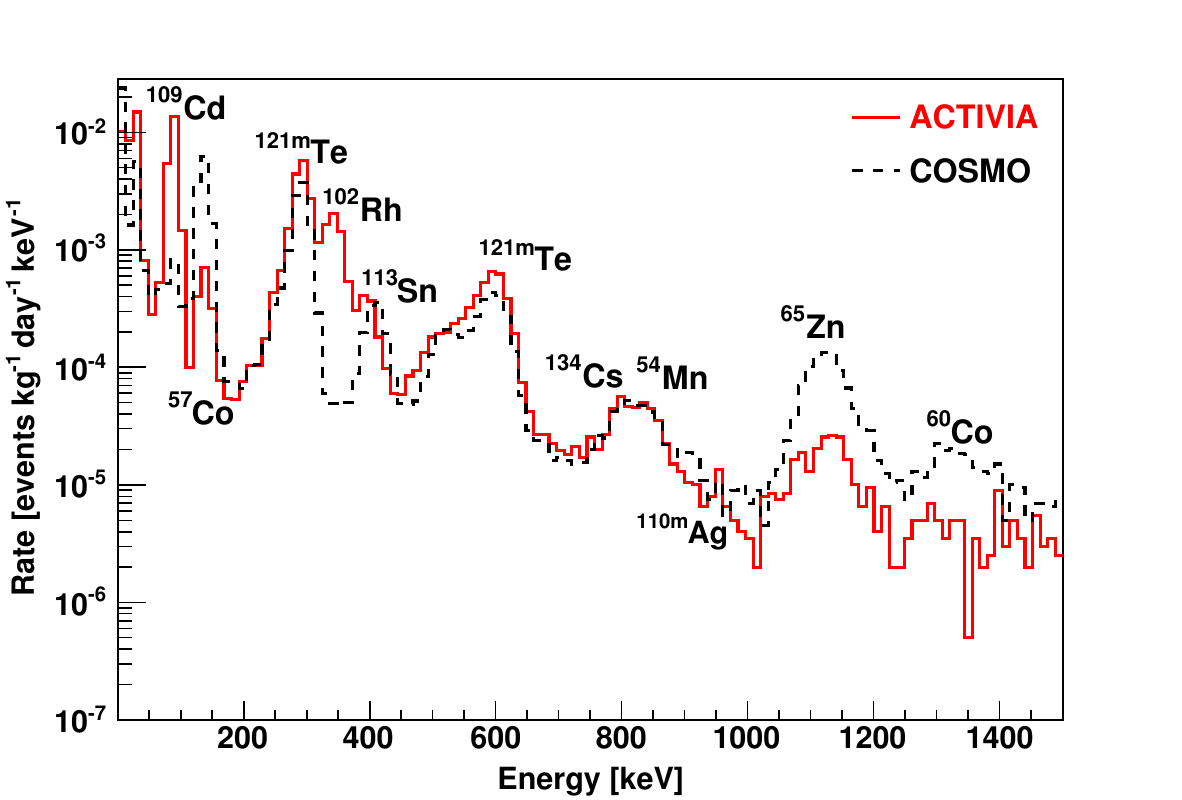}
\caption{(Color online) Predicted background from the cosmogenic xenon activation. The background from natural radioactivity is not included. The production rates have been calculated with ACTIVIA~\cite{ACTIVIA} and COSMO~\cite{COSMO} packages, the radioactive decays have been simulated in the LXe target with GEANT4, assuming 1~year of activation on the earth's surface and 2~years of cooldown time underground. Radioactive isotopes with half-lives larger than 100~days have been selected. Only the most prominent peaks are labeled. The measured energy resolution is taken into account.}
\label{figCosmogenics}
\end{center}
\end{figure}

Another discrepancy between the measured and Monte Carlo background spectra is present in the energy region 700-1100~keV. The cosmogenic activation of natural xenon during storage at Earth's surface might be responsible for this discrepancy. This has been studied assuming 1~year of activation and 2~years of cooldown time with the ACTIVIA~\cite{ACTIVIA} and COSMO~\cite{COSMO} simulation packages. Both use semi-empirical formulae~\cite{SilberbergTsao} to estimate the cross-sections of nuclear processes. The calculated production rates have been used as an input for a Monte Carlo simulations of the radioactive decays with GEANT4. The simulated energy spectra of the cosmogenic background are shown in Figure~\ref{figCosmogenics}. The production rates predicted by the packages differ by one order of magnitude and more, and the discrepancy between 700-1100~keV in Figure~\ref{figDataMC} cannot be explained without destroying the remarkable agreement in other energy ranges. 
A similar calculation has been performed for a natural xenon target in Ref.~\cite{CosmogenicProduction} using the TALYS code \cite{TALYS}, and the published results do not agree with either ACTIVIA or COSMO in terms of isotopes produced by cosmogenic activation and their production yields. 

The theoretical spectrum of the 2$\nu$ double beta decay of $^{136}$Xe is also shown in Figure~\ref{figDataMC}, assuming the half-life limit of (2.11$\pm$0.04$\pm$0.021)$\times$10$^{21}$~years~\cite{EXO}. Its contribution does not change the total background spectrum significantly, thus it can be concluded that the observed discrepancy around 700-1100~keV cannot be explained by this potential background source. The predicted energy averaged background rate from the 2$\nu~\beta\beta$ decay of $^{136}$Xe is at the level of 10$^{-6}$~events$\cdot$kg$^{-1}\cdot$day$^{-1}\cdot$keV$^{-1}$ below 100~keV, three orders of magnitude lower than the background from other components.

\section{Conclusions}
\label{secConclusions}
An extensive study to predict the electronic recoil background of the XENON100 experiment has been performed . The study is based on Monte Carlo simulations with GEANT4 using a detailed mass model of the detector and its shield, and the measured radioactivity values of all relevant detector components. 

The design goal of XENON100, to gain a factor of 100 reduction in background rate compared to XENON10 (0.6~events$\cdot$kg$^{-1}\cdot$day$^{-1}\cdot$keV$^{-1}$~\cite{xe10-independent}), has been achieved. This has been possible thanks to a selection of all detector materials, an innovative design of the cryogenic system, the use of an active LXe veto, and an improved passive shield.

The predicted upper limit on the average single scatter electronic recoil rate in the energy region below 100~keV, without an active veto veto is 17.3$\times$10$^{-3}$ (10.8$\times$10$^{-3}$) events$\cdot$kg$^{-1}\cdot$day$^{-1}\cdot$keV$^{-1}$ for 40~kg (30~kg) fiducial mass (Table~\ref{tab:summaryElectronRecoils}). By applying a veto cut with an average energy threshold of 100~keV, these rates are reduced to 7.0$\times$10$^{-3}$ (5.3$\times$10$^{-3}$)~events$\cdot$kg$^{-1}\cdot$day$^{-1}\cdot$keV$^{-1}$ for 40~kg (30~kg) fiducial mass. The discrimination between electron and nuclear recoils based on the ratio of proportional to primary scintillation light (S2/S1) is not considered in this paper, and provides a further background reduction of~$>$99\%.

From the good agreement between Monte Carlo simulation and measured data, as shown in Figures~\ref{figDataMC} and \ref{figDataMCzoom}, and the predicted background rates from Table~\ref{tab:summaryElectronRecoils}, it can be concluded that the electronic recoil background in the XENON100 experiment during the commissioning run in Fall 2009~\cite{xe100-independent} is dominated by the natural radioactivity in the detector materials. With an optimized fiducial volume cut and an active veto cut, the background rate in the energy region of interest can be reduced down to a level where radioactive $^{85}$Kr in LXe starts to dominate.

The results of the present work are not only important for understanding the electromagnetic background in the XENON100 experiment and the validation of the background model, but can be also useful for the design of next-generation detectors for dark matter searches, such as XENON1T~\cite{xenon1t} or DARWIN~\cite{darwin}.

\section{Acknowledgements}
This work has been supported by the National Science Foundation Grants No.~PHY-03-02646 and PHY-04-00596, the Department of Energy under Contract No.~DE-FG02-91ER40688, the CAREER Grant No.~PHY-0542066, the Swiss National Foundation SNF Grant No.~20-118119, the Volkswagen Foundation, and the FCT Grant No.~PTDC/FIS/100474/2008.

\end {document}